\documentclass{optica-article}

\journal{opticajournal} % for journals or Optica Open

\articletype{Research Article}

\begin{document}

\title{Chaos with Gaussian invariant distribution by quantum-noise random phase feedback}

\author{Yanqiang Guo,\authormark{1,2} Haifeng Li,\authormark{1,2} Yingqi Wang,\authormark{1} Xiangyu Meng,\authormark{1} Tong Zhao,\authormark{1} and Xiaomin Guo\authormark{1,*}}

\address{\authormark{1}Key Laboratory of Advanced Transducers and Intelligent Control System, Ministry of Education, College of Physics, Taiyuan University of Technology, Taiyuan 030024, China\\
\authormark{2}State Key Laboratory of Cryptology, Beijing 100878, China\\}
%\authormark{3}Currently with the Department of Electronic Journals, Optica Publishing Group, 2010 Massachusetts Avenue NW, Washington, DC 20036, USA}

\email{\authormark{*}guoxiaomin@tyut.edu.cn} %% email address is required; see note below about the corresponding author designation

% use {asbstract*} to suppress the copyright line. Copyright information will be added in production

\begin{abstract*}
We experimentally present a random phase feedback based on quantum noise to generate a chaotic laser with Gaussian invariant distribution. The quantum noise from vacuum fluctuations is acquired by balanced homodyne detection and injected into a phase modulator to form a random phase feedback. An optical switch using high-speed intensity modulator is employed to reset the chaotic states repeatedly and the time evolutions of intensity statistical distributions of the chaotic states stemming from the initial noise are measured. By the quantum-noise random phase feedback, the transient intensity distributions of the chaotic outputs are improved from asymmetric invariant distributions to Gaussian invariant distributions, and the Gaussian invariant distribution indicates a randomly perturbed dynamical transition from microscopic initial noise to macroscopic stochastic fluctuation. The effects of phase feedback bandwidth and modulation depth on the invariant distributions are investigated experimentally. The chaotic time-delay signature and mean permutation entropy are suppressed to 0.036 and enhanced to 0.999 using the random phase feedback, respectively. The high-quality chaotic laser with Gaussian invariant distribution can be a desired random source for ultrafast random number generation and secure communication.

\end{abstract*}

%%%%%%%%%%%%%%%%%%%%%%%%%%  body  %%%%%%%%%%%%%%%%%%%%%%%%%%
\section{Introduction}
Chaotic semiconductor laser with time-delayed feedback is an important system for generating high-dimensional complex outputs due to its excellent nonlinear dynamical characteristics such as noise-like, broadband power spectrum and sensitivity to initial conditions \cite{Soriano13,Chembo19}. It is widely used in secure communication \cite{Argyris05,Hong08,Jiang19,Xue16,Wu13}, high-speed physical random number generation \cite{Uchida08,Wang17,Virte14,Guo19,Tang15,Wu2012}, optical sensing \cite{Ma15}, chaotic radar \cite{Lin04}, optical time-domain reflectometer \cite{Wang08}, precision measurement and control of photon statistics \cite{Guo18,Guo22} and so on. Although the external cavity feedback induces high dimensional chaos, it also brings the detrimental period which deteriorates the security of chaotic laser and limits its applications. Meanwhile, dynamic instability of chaos can be characterized by time evolution of probability density distribution of trajectory ensemble, but the probability distribution always appears as a complex deformation or asymmetry in the transient process. Although it eventually tends towards an invariant distribution, it cannot reach a normal random distribution. The asymmetry distribution indicates that the randomness originating from microscopic noise is reduced in the chaotic amplification process. It is of great importance to improve the transient invariant distribution for the preparation of high-security and strong-randomness chaotic laser.

In recent years, many post-processing methods have been developed to improve the randomness of the chaotic outputs, such as dual-path feedback \cite{Wu09}, fiber grating feedback \cite{Zhong17,Xu17}, frequency detuned feedback \cite{Li2015}, analog-digital hybrid feedback \cite{Cheng18}, optical injection \cite{Li12,Li15,Xiang16,Wu12}, fiber propagation \cite{Li18}, exclusive-OR operation \cite{Uchida08}, the m-least significant bit selection \cite{Kanter10}, quantum noise injection \cite{Guo2021}, Frequency-Band Extractor \cite{Guo20,Guo21}, optoelectronic heterodyne \cite{Wang13} and self-phase-modulated feedback \cite{Jiang18,Zhao19,Xiang14,Mu17,Ma20}. The time delay period of the chaotic dynamics is effectively suppressed, but there is no detailed and direct experimental observation of the noise enhancement effect on the transient and long-term statistical properties of full-developed chaotic process. The evolutionary process of the noise effect and the random signal extraction in an effective way remains to be explored \cite{Sunada12}. Moreover, due to a new frequency component introduced by phase modulation, feedback mechanism of random phase of microscopic noise and its influence on macroscopic chaotic outputs need to be further studied.

In this work, we propose an experimental scheme of quantum-noise random phase feedback (QNRPF) to prepare a high-performance chaotic laser with Gaussian invariant distribution. The quantum noise is extracted by balanced homodyne detection and used to build a random phase feedback loop. We also employ a high-speed optical switch to reset the laser dynamics to an original state repeatedly and measure the time evolution of the chaos statistical distributions. The transient invariant distribution of chaotic output is measured and improved by the QNRPF, and the improved statistical skewness of the long-term intensity distribution is observed. The invariant distributions for various phase feedback bandwidths and modulation depths are studied experimentally. The QNRPF suppresses the chaotic time-delay signature (TDS) to the noise level.  High-resolution map of phase feedback bandwidth and modulation depth versus TDS are obtained. In addition, the improved mean permutation entropy of the chaotic outputs is revealed. The technique plays an important role in random number generation and secure communication.

\section{Experimental setup}

\begin{figure}[htbp]
\centering\includegraphics[width=12cm]{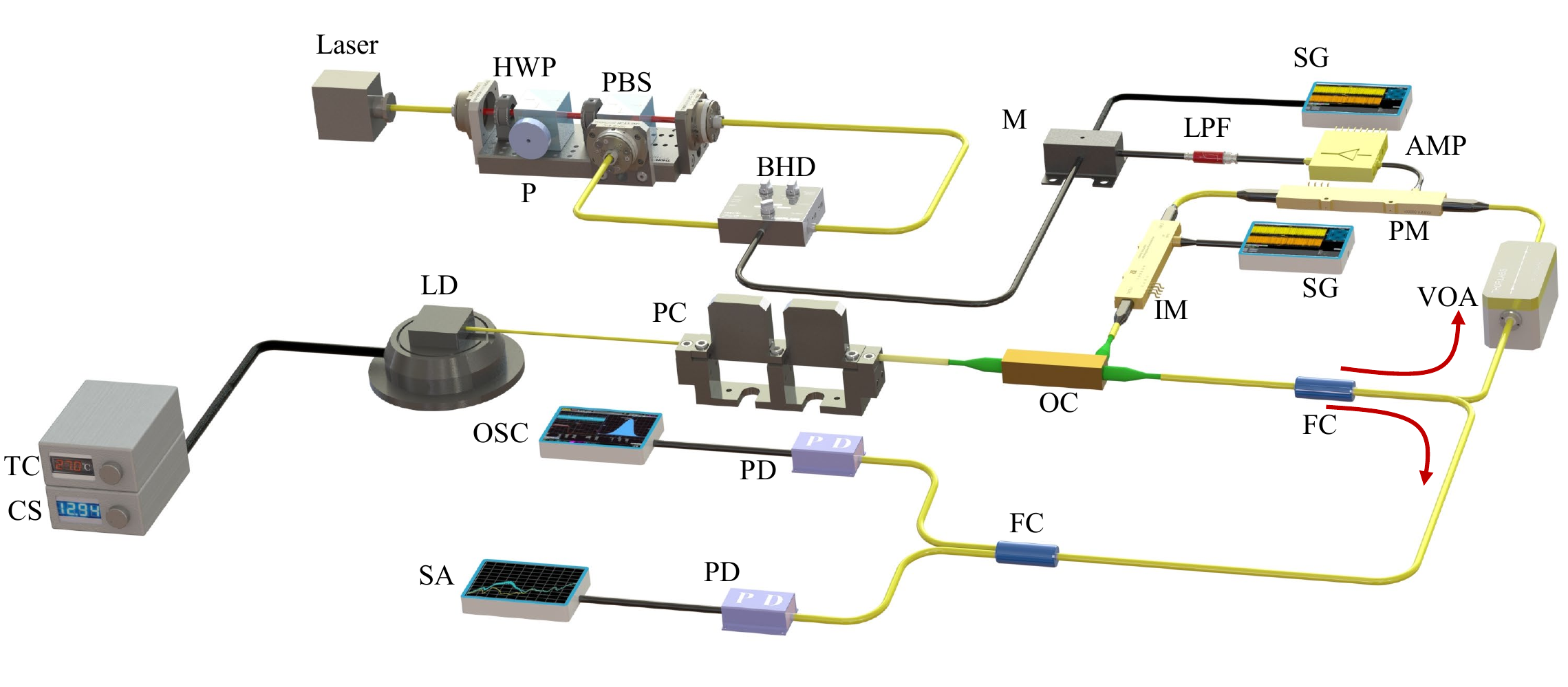}
\caption{Scheme of experimental setup. HWP: half-wave plate; P: power meter; PBS: polarizing beam splitter; BHD: balanced homodyne detector; SG: signal generator; LPF: low -pass filter; AMP: wideband amplifier; IM: intensity modulator; PM: phase modulator; TC: temperature controller; CS: current source; LD, distributed feedback laser diode; PC: polarization controller; OC: optical circulator; VOA: variable optical attenuator; FC: fiber coupler; PD: photodetector; OSC: oscilloscope; SA: spectrum analyzer.}
\label{fig1}
\end{figure}

The experimental setup of the scheme is shown in Fig. \ref{fig1}. An external cavity distributed feedback semiconductor laser (LD) is stably controlled by a low noise temperature controller (TC) with an accuracy of 0.01 ℃ and a current source (CS) with an accuracy of 0.1 mA. The LD threshold current is 6 mA. The polarization of the LD is adjusted by a polarization controller (PC), and then the output passes through an optical circulator (OC) and a 20:80 fiber coupler (FC). The 80$\%$ of optical output is fed back to the LD via a variable optical attenuator (VOA, accuracy 0.01 dB), phase modulator (PM, Photline MX-LN-10) and intensity modulator (IM, Ixblue MPZ-LN-10), forming an external cavity feedback loop. The chaotic feedback strength is controlled by the VOA. The half wave voltage of PM is 4 V corresponding to the peak phase shift of $\pi$ , and the half wave voltage of IM is 3.8 V. The PM is driven by a quantum shot noise signal and the IM is driven by a square-wave pulse signal from a signal generator (SG). The IM operates stably in the linear range using a modulator bias controller, and a high-speed optical switch is built by the IM square modulation. Quantum shot noise is extracted by a balanced homodyne detector (BHD, Thorlabs PDB480C) and amplified to macro level via local oscillator (LO) and electrical gains. A 1550 nm single-mode laser (Laser) is used as the LO. One group of half wave plate (HWP) and polarizing beam splitter (PBS) is combined to ensure that the LO interferes with quantum vacuum state. Another group of HWP and PBS is employed as accurate 50:50 beam splitters to divide the input beam equally, and the transmitted and reflected outputs are coupled separately into two single-mode optical fibers by two triplet-lens couplers. The measured quantum shot noise is mixed with a radio frequency signal generated by a SG, and filtered by a low-pass filter (LPF) with different bandwidths. The intensity of the filtered quantum noise is precisely amplified and adjusted by a wideband amplifier (AMP) and then the output is injected into the PM in the chaotic external cavity feedback loop. It forms a random phase feedback by the quantum noise injection. The final outputs are detected by two 50 GHz bandwidth photodetectors (PD, Finisar, XPDV212ORA-VF-FP). The power spectrum and time series are acquired by a 26.5 GHz spectrum analyzer (SA, Agilent N9020A) and a real-time oscilloscope with a bandwidth of 36 GHz  (OSC, Lecroy LabMaster10-36Zi).

\section{Experimental results}

\subsection{High-speed optical switch for resetting chaotic state}
In order to investigate the amplification and evolution process of chaotic dynamics, we first build a high-speed optical switch using the IM to reset the chaotic state repeatedly. The time evolution of chaotic signals originating from initial noise can be obtained, which indicates the dynamic behavior and complex statistical properties of chaotic laser. 
\begin{figure}[htbp]
\centering\includegraphics[width=11cm]{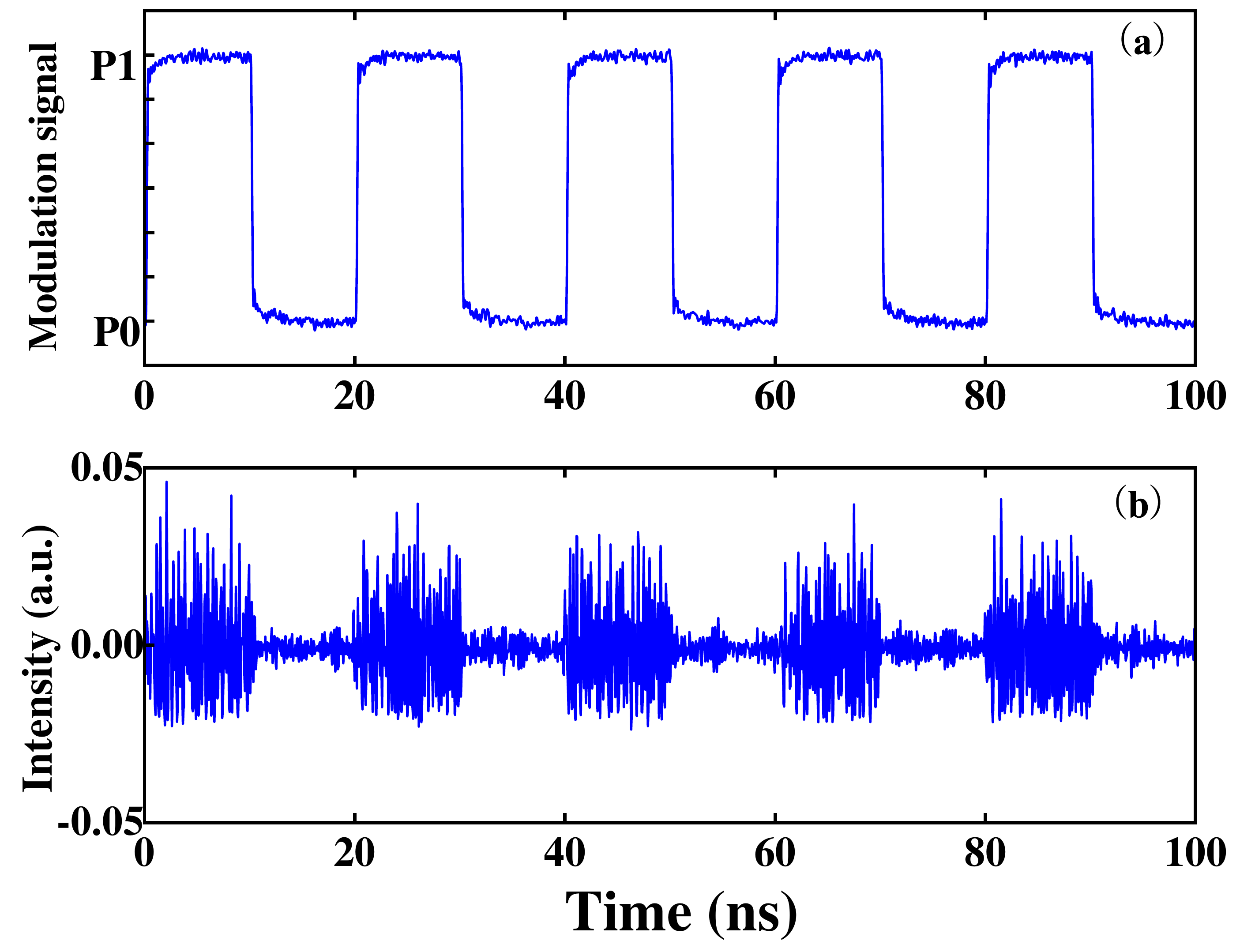}
\caption{Measured chaotic laser intensity using optical switching. (a) measured square-wave modulation signal; (b) time series of chaotic laser when the optical switch is on and off.}
\label{fig2}
\end{figure}
The detailed signal variation of the optical switch is shown in Fig. \ref{fig2}, where the output signal alternates between stable coherent state and chaotic state. In Fig. \ref{fig2}(a) the open voltage P1 of the optical switch is optimized to 3 V for obtaining the reset chaotic state. When the low voltage P0 is closed to 0 V, the optical feedback is switched off and it outputs a stable coherent laser. The rise time of the on-off operation is on the order of hector-picosecond to achieve fast switching of different dynamic states. The switching period is 20 ns, which is long enough to avoid other unstable factors caused by modulation. In Fig. \ref{fig2}(b), it can be seen that the measured signal quickly transitions from the stable laser state to the chaotic state by controlling the optical feedback on and off. This phenomenon can be explained by the dynamical behavior of chaotic laser system. In the unstable state, the system is located near the chaotic attractor, meaning that initial noise perturbations will quickly bring the system towards the attractor. As time progresses, the chaotic state drifts around the attractor, exhibiting complex time-evolution.

\begin{figure}[htbp]
\centering\includegraphics[width=12cm]{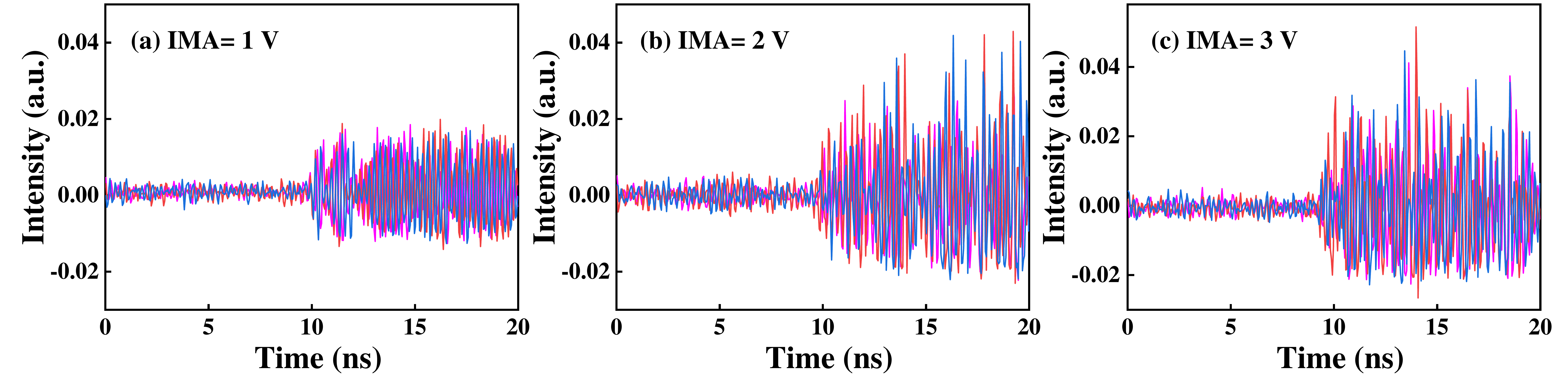}
\caption{Time series evolutions of chaotic intensity starting from the initial state for different IMAs of optical switch. The optical switch is turned on at the time of 10 ns. The voltages of IMA are (a) 1 V, (b) 2 V, and (c) 3 V, respectively.}
\label{fig3}
\end{figure}

Then, we measured the chaotic intensity evolutions from stable state to transient state for different modulation amplitudes. Figure \ref{fig3} shows the transient process starting from the initial stable state. Figures. \ref{fig3}(a)-\ref{fig3}(b) show different time series of the chaotic laser for three intensity modulation amplitudes (IMAs) of 1 V, 2 V and 3 V. The time t from the stable laser to the dynamic amplification is about 10 ns. When the t < 10 ns, the optical feedback is off and the LD laser outputs a stable coherent state. When the t > 10 ns, the initial noise amplified by chaotic dynamics leads to trajectory separation of intensity fluctuations, and the output time series exhibit complex dynamic behavior.

\begin{figure}[htbp]
\centering\includegraphics[width=12cm]{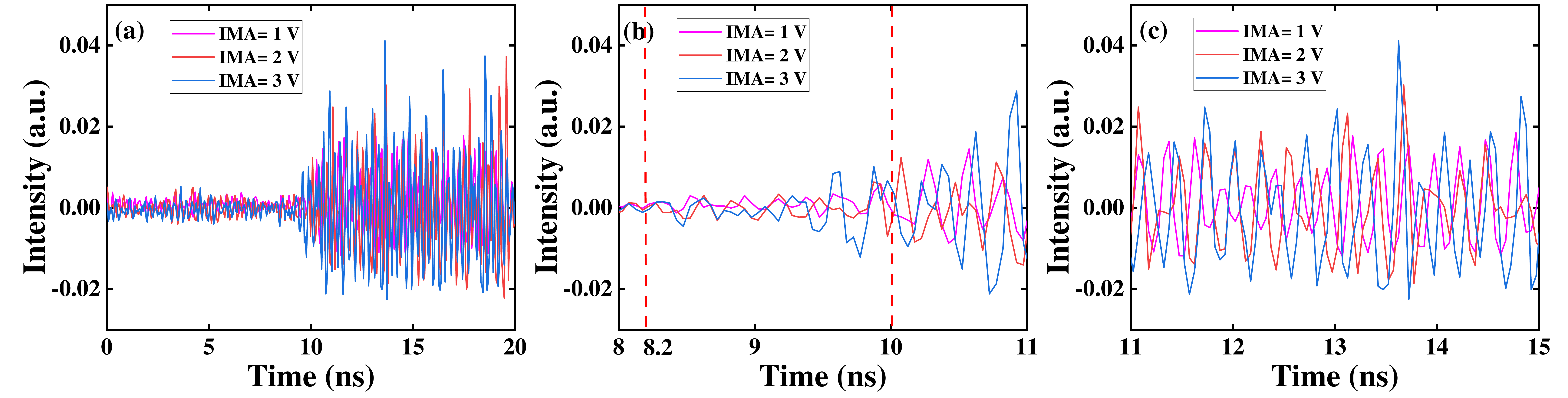}
\caption{(a) Three time series of chaotic intensities starting from the initial state for IMAs of 1 V, 2 V, and 3 V. (b) Evolution process of the intensity trajectories from 8 ns to 11 ns. (c) Evolution process of the intensity trajectories from 11 ns to 15 ns.}
\label{fig4}
\end{figure}

In order to further explore the amplification and evolution process of dynamic instability caused by inherent laser noise, we measure and compare the output intensity evolution from stable noise state to transient state in three IMAs of 1 V, 2 V, and 3 V, as show in Fig. \ref{fig4}. As the IMA increases, the speed of intensity trajectory separation accelerates and the dynamic amplification also becomes more rapid. Specifically, as shown in Fig. \ref{fig4}(b), for different intensity modulation amplitudes of the high-speed optical switch, 
the intensity trajectory starts to separate at 8.2 ns. The fully separated state of chaotic amplification has been achieved within approximately 2 ns. In the subsequent time period, the separations of the intensity trajectories become more pronounced, as shown in Fig. \ref{fig4}(c). These results indicate that in the chaotic laser system, microscopic initial-noise perturbations can lead to significant intensity trajectory separation. As the IMA increases, the output amplitude of the chaotic fluctuations increases. The initial perturbations induce fast dynamic amplification and trajectory separation.

\subsection{Gaussian invariant distribution with QNRPF}

\begin{figure}[htbp]
\centering\includegraphics[width=12.6cm]{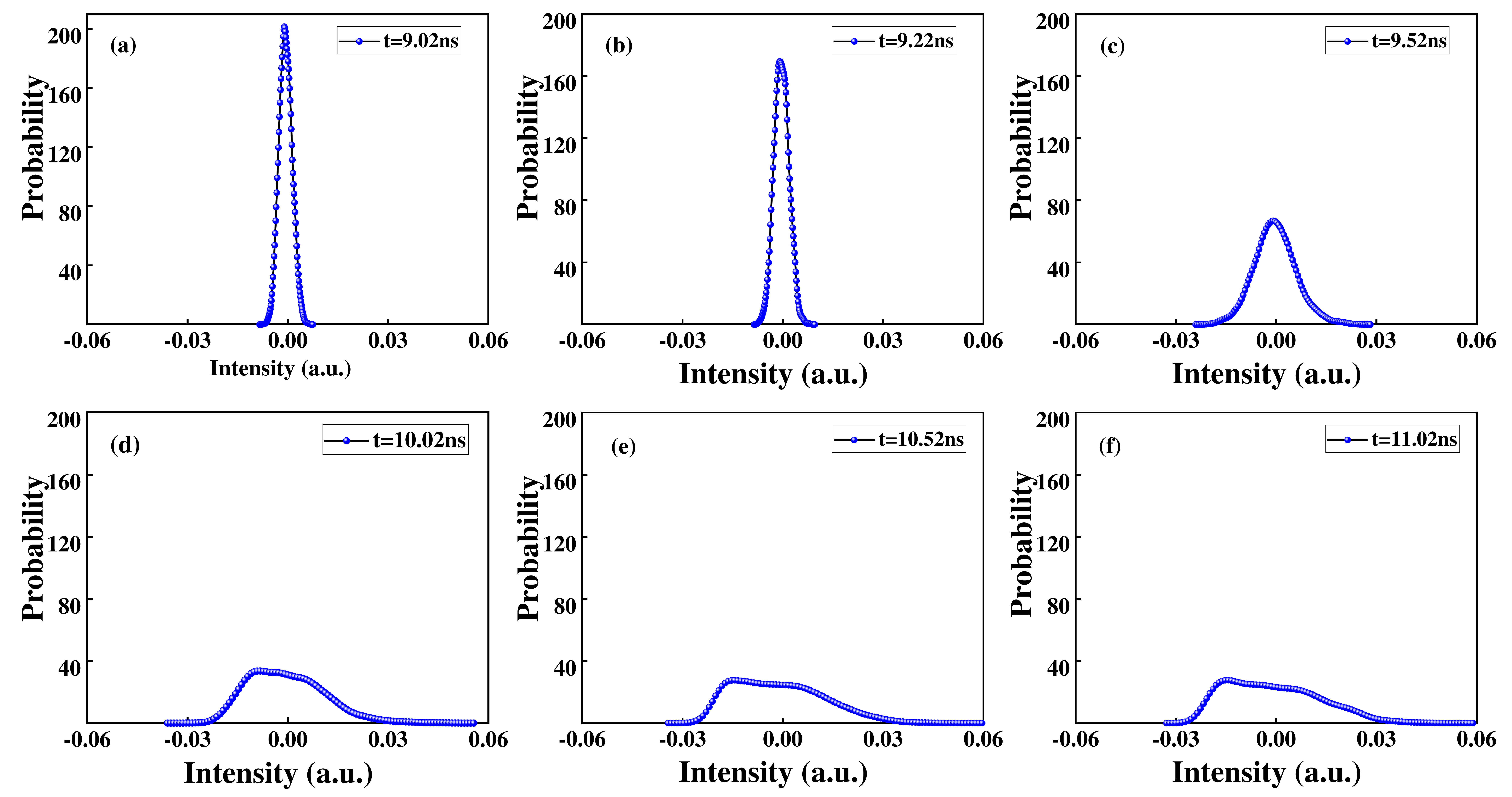}
\caption{Transient intensity distributions of chaotic outputs without QNRPF, when the IMA is 3 V.}
\label{fig5}
\end{figure}

The nonlinear dynamical and statistical characteristics of chaotic laser can be well revealed by transient and long-term probability density distribution. To illustrate the chaotic system randomly perturbed by initial noise, we measure intensity distributions of chaotic transient states without QNRPF. The optical switch is repeatedly switched with the IMA of 3 V and the transient distribution of the chaotic outputs is obtained from the fixed-time intensities of  2×$10^{4}$ time series. Figures \ref{fig5}(a)-\ref{fig5}(f) show the transient probability distribution of the chaotic intensity changes with the transient time. As the transient time increases above 10 ns, the transient intensity distribution of chaotic signals gradually converges to an invariant probability density distribution. However, the invariant distribution without QNRPF presents asymmetry, and it indicates that the output signals do not maintain the randomness of the initial noise although the chaotic system is randomly perturbed.

\begin{figure}[htbp]
\centering\includegraphics[width=12.6cm]{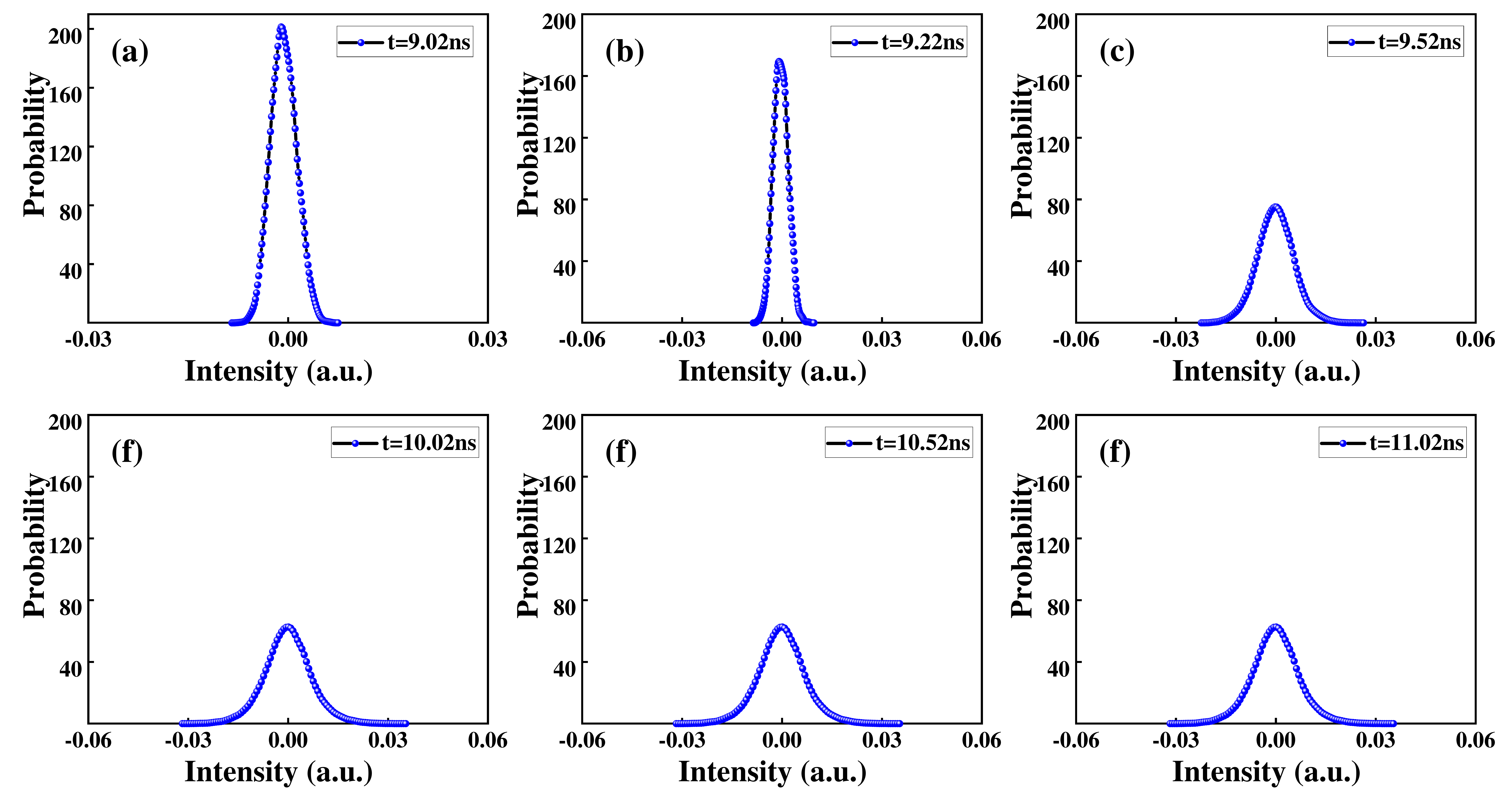}
\caption{Transient intensity distributions of chaotic outputs with QNRPF. The parameters of intensity modulation are the same as those in Fig. \ref{fig5}, and the phase modulation bandwidth and depth of the QNRPF are 900 MHz and 2.5, respectively.}
\label{fig6}
\end{figure}

To retain the initial randomness and prevent the detriment of external-cavity period, we develop a random phase feedback using quantum noise injection to improve the transient invariant distribution and the randomness of the output signals. The transient distributions of the chaotic intensities are measured with the QNRPF and repeated optical switching. In the QNRPF, the phase feedback bandwidth of quantum noise injection is 900 MHz and the phase modulation depth is 2.5, which corresponds to a peak phase shift of 2.5 $\pi$. The half wave voltage of the PM is 4 V, and the phase modulation depth is controlled by adjusting the peak-to-peak amplitude of the PM driving signal. The transient intensity distributions of the chaotic outputs are effectively improved by the random phase perturbation. Figs. \ref{fig6}(a)-\ref{fig6}(f) show that the random phase feedback induces the transient probability distribution of chaotic signals to converge to a Gaussian invariant distribution, which is symmetrically distributed with no skew. The results indicate that the QNRPF enhances the randomly perturbed effect on the dynamic evolution process and the trajectories of the same initial state behaves random separations.

\begin{figure}[htbp]
\centering\includegraphics[width=12cm]{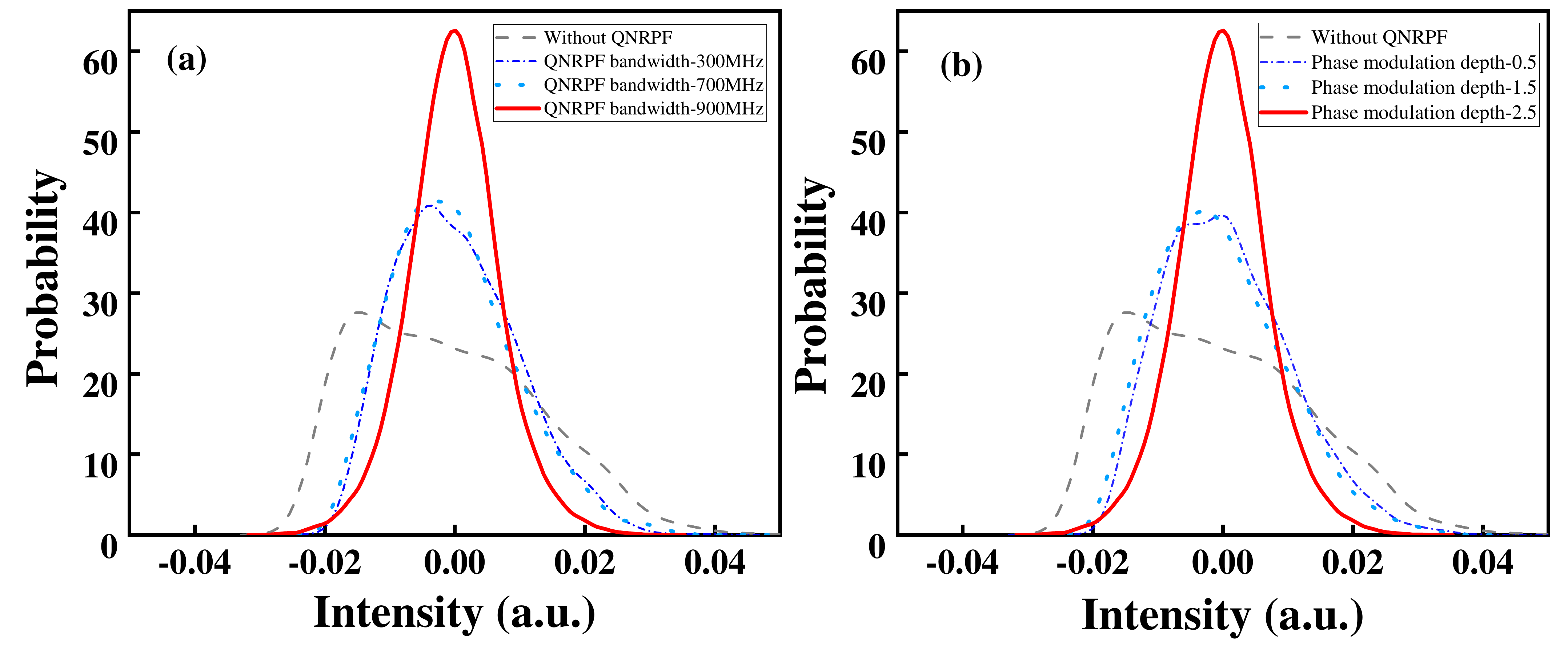}
\caption{Transient invariant distribution of chaotic intensity at the time $t=11.02$ ns. (a)When the phase modulation depth is 2.5, the invariant distribution is without and with QNRPF for various QNRPF bandwidths of 300 MHz, 700 MHz and 900 MHz; (b)When the QNRPF bandwidth is 900 MHz, the invariant distribution is without and with QNRPF for various phase modulation depths of 0.5, 1.5 and 2.5.}
\label{fig7}
\end{figure}

To further investigate the effects of the QNRPF bandwidth and phase modulation depth, we measure the transient intensity distributions for various phase feedback bandwidths and modulation depths. At the transient time of 11.02 ns, the invariant intensity distributions are recorded for various QNRPF bandwidths and the phase modulation depth of 2.5, as shown in Fig. \ref{fig7}(a). It can be seen that the transient intensity distribution without the QNRPF shows an asymmetric distribution with a large width. As the phase feedback bandwidth increases from 300 MHz to 900 MHz, the invariant distribution gradually converges to a symmetric Gaussian distribution. The results indicate that the effect of the random phase perturbation increases as the bandwidth of the random phase feedback increases, 
and the high-bandwidth random phase feedback contributes to enhancing the randomness of the chaotic output signals. Figure \ref{fig7}(b) shows the effect of phase modulation depth on the transient invariant distribution of chaotic intensity when the QNRPF bandwidth is 900 MHz. As the phase modulation depth increases from 0.5 to 2.5, the invariant distribution changes from the initial asymmetric shape to the symmetric Gaussian shape. By increasing the bandwidth of the random phase feedback and phase modulation depth, the invariant distribution of chaotic transient state is improved and the randomly perturbed dynamics of the chaotic laser is enhanced.

\subsection{Time-delay signature suppression and mean permutation entropy enhancement with QNRPF}

\begin{figure}[htbp]
\centering\includegraphics[width=12.6cm]{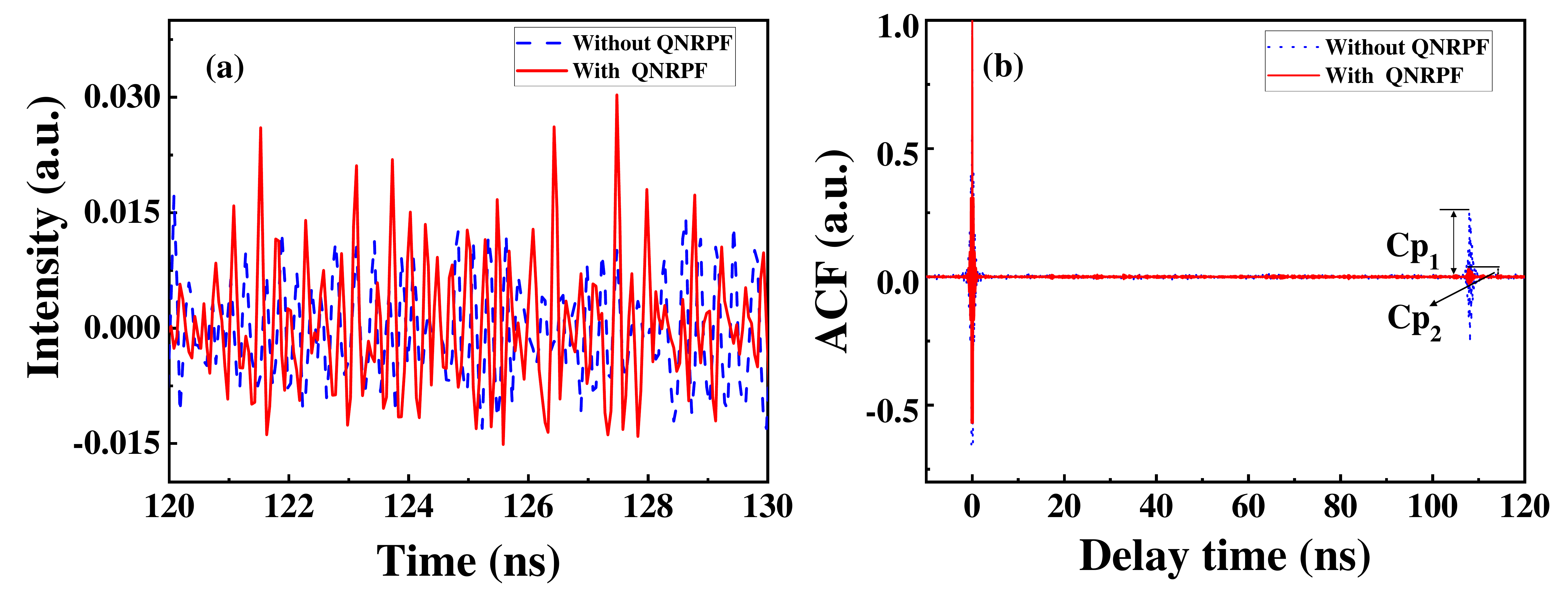}
\caption{(a) Time series and (b) ACF of chaotic laser with and without the QNRPF. The QNRPF bandwidth is 900 MHz and its phase modulation depth is 2.5.}
\label{fig8}
\end{figure}

In order to further verify the complexity and entropy of the output signals while improving the invariant distribution, we study chaotic time-delay signature (TDS) and mean permutation entropy (MPE) with the QNRPF. In our experiment, the peak value $ C_{p} $ of autocorrelation function (ACF) at the feedback delay time is used to quantify the TDS of chaotic laser. The ACF that measures the time series between a signal and its time-delay version is defined as follows:
\begin{equation}
C(\Delta t) =\frac{\left\langle\left[I(t+\Delta t)-\left\langle I(t+\Delta t)\right\rangle\right]\left[I(t)-\left\langle I(t)\right\rangle\right]\right\rangle}{\sqrt{\left\langle\left[I(t+\Delta t)-\left\langle I(t+\Delta t)\right\rangle\right]^2\right\rangle\left\langle\left[I(t)-\left\langle I(t)\right\rangle\right]^2\right\rangle}},  \\
\end{equation}
where \textit{I(t)} represents the intensity of the chaotic laser signal, $\Delta t$ represents the delay time, and <·> represents the average of time. Figure \ref{fig8} shows the time series and time-delay signature of the chaotic laser for the QNRPF bandwidth of 900 MHz and phase modulation depth of 2.5. The red solid line in Fig. \ref{fig8}(a) indicates the time series with the QNRPF, and the mean and peak-to-peak values of the long-term fluctuating amplitudes are little changed compared to those without the QNRPF. It is worth noting that the TDS is effectively suppressed using the QNRPF and the peak value of the TDS falls from the $ C_p{}_1 $ of 0.259 to the $ C_p{}_2 $ of 0.036, which reaches the noise level. The periodicity of the external-cavity time delay is significantly reduced by the QNRPF.

\begin{figure}[htbp]
\centering\includegraphics[width=12.6cm]{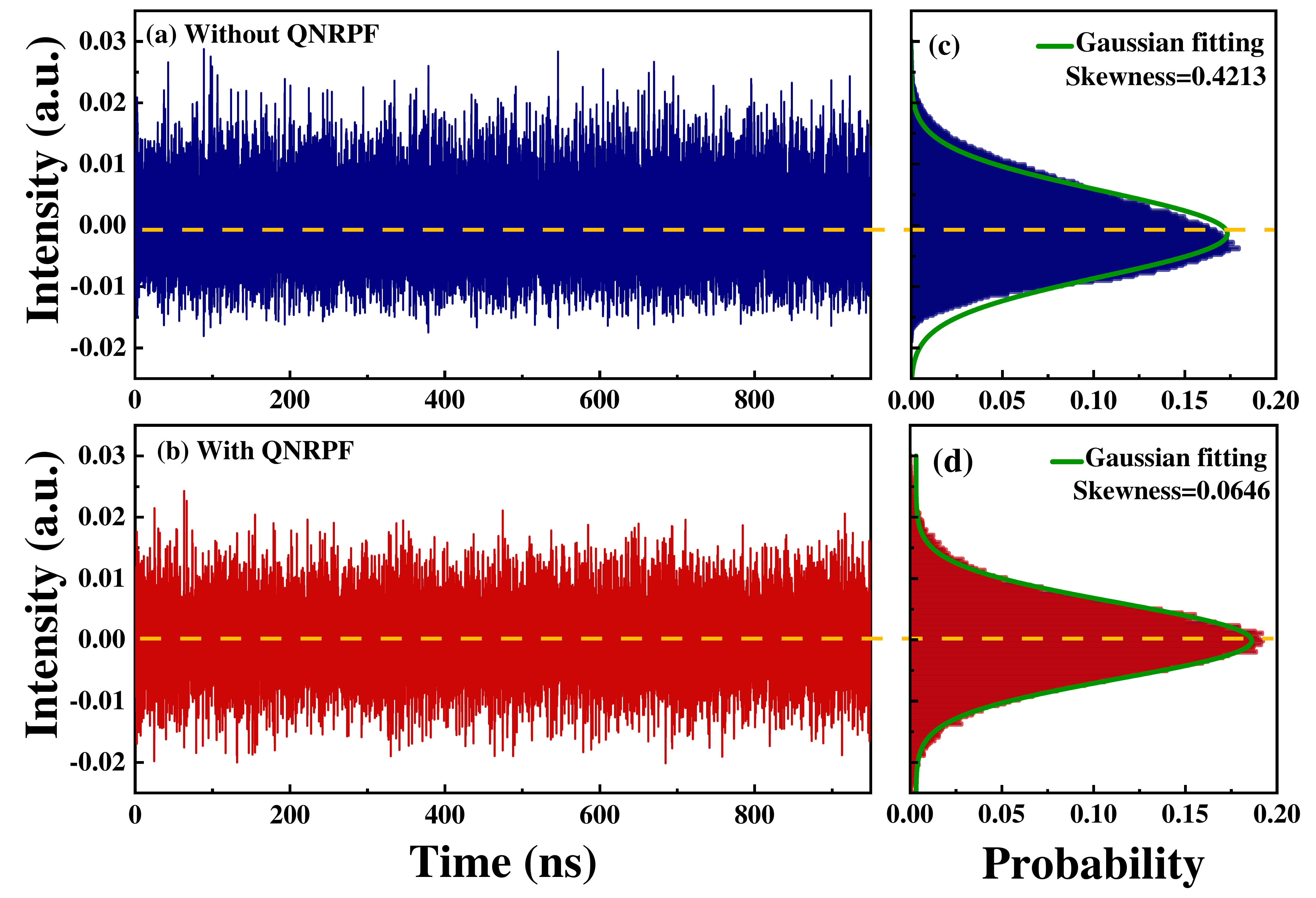}
\caption{Chaotic time series (a) without QNRPF and (b) with QNRPF, corresponding to the long-term intensity distributions (c) and (d). The solid line represents the Gaussian fitting and the dashed line represents the mean intensity.}
\label{fig9}
\end{figure}

In addition to suppressing the TDS, the QNRPF can also improve the statistical skewness of the long-term chaotic intensity distribution. Figures \ref{fig9}(a)-\ref{fig9}(b) show the $10^{6}$ recorded data for each time series without and with the QNRPF, respectively. The corresponding long-term intensity distributions are shown in Fig. \ref{fig9}(c)-\ref{fig9}(d), where the solid line is the Gaussian fitting and the dashed line is the mean value of intensity. The long-term intensity distribution of the time series without the QNRPF is skewed to one side and the skewness is 0.4213, due to the time delay period. The long-term intensity distribution with the QNRPF is significantly improved and its skewness is reduced to 0.0646. The phase random feedback reduces the skewness of the long-term time series distribution by an order of magnitude. The improvement of the time series distribution is beneficial to the randomness extraction from chaotic laser and chaos-based secure communication.

\begin{figure}[htbp]
\centering\includegraphics[width=12.6cm]{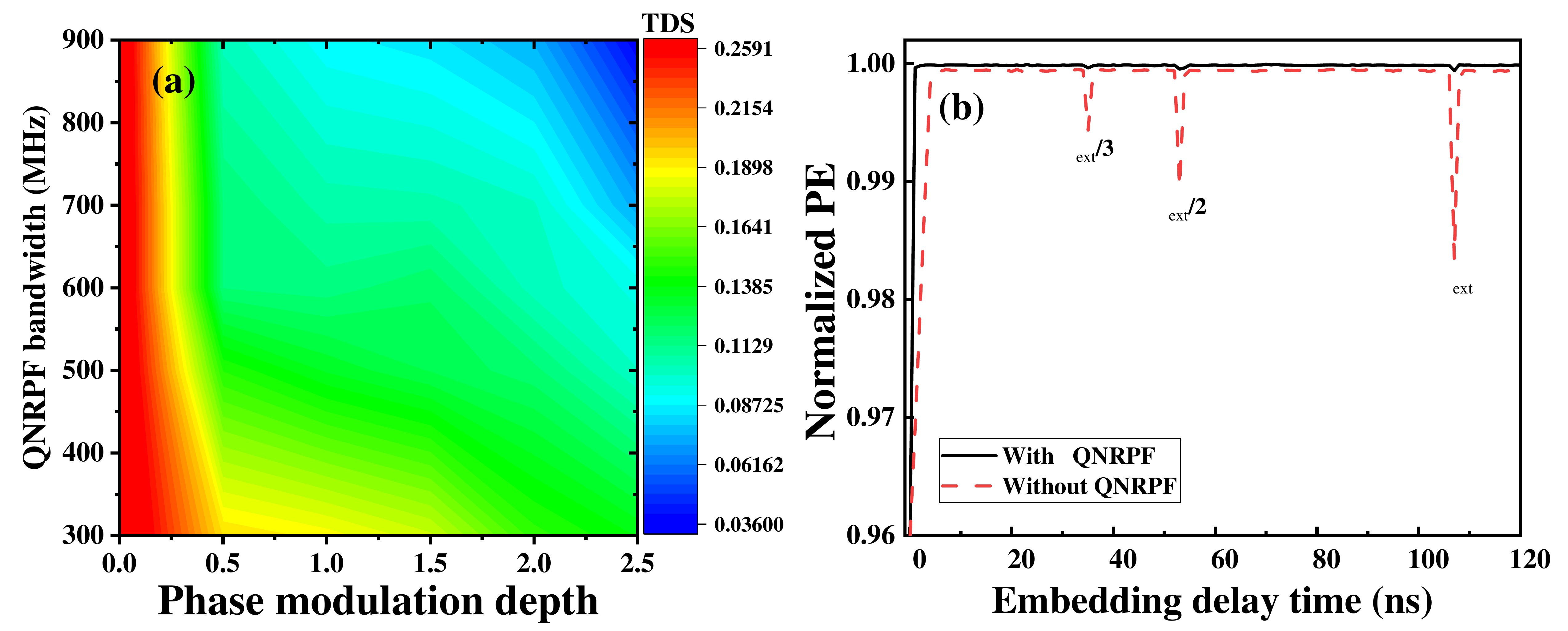}
\caption{(a) Measured map of TDS variation with phase modulation depth and QNRPF bandwidth. (b) Normalized PE without QNRPF (red dashed line) and with QNRPF (black solid line) versus embedding delay time.}
\label{fig10}
\end{figure}

For clarity, the effects of QNRPF bandwidths and phase modulation depths on the TDS are experimentally investigated. High-resolution map of the TDS over wide range of the bandwidth-depth parameter space is further measured in detail, as shown in Fig. \ref{fig10}(a). When the phase modulation depth is less than 0.5, the TDS of the chaotic laser is not well suppressed even with increasing the QNRPF bandwidth. As the phase modulation depth and the QNRPF bandwidth gradually increase, it can be clearly seen that the TDS is effectively suppressed. 

Meanwhile, we employ mean permutation entropy (MPE) at the feedback delay time $ \tau_{ext} , \tau_{ext}/2 , \tau_{ext}/3 $ to quantify the complexity of the chaotic signal. Permutation entropy is first introduced by Bandt and Pompe \cite{Bandt02} to quantify the unpredictability of time series and has strong robustness against environmental noise. A full arrangement method is to count the occurrence number of various permutations and calculate their relative frequencies to obtain the probability. The permutation entropy values of 0 and 1 represent completely predictable and stochastic process respectively. Embedding dimension $ \textit{d} $ can be selected between 3 and 7 in experiment. When the embedding dimension $ \textit{d} $ = 4, the MPE including the three main TDS peaks are defined as follows: 
\begin{equation}
h_d^{MPE}=\frac{h_d^Text+h_d^Text/2+h_d^Text/3}{3},
\end{equation}
where $ h_d^{\tau_{ext}} , h_d^{\tau_{ext/2}} , h_d^{\tau_{ext/3}} $ represent the normalized PE values at the delay time of $ \tau_{ext} , \tau_{ext}/2 , $ $\tau_{ext}/3 $. The MPE can accurately evaluate the entropy changes of the chaotic laser system in physical process. Figure \ref{fig10}(b) shows the permutation entropy values at the three delay times apparently decrease without the QNRPF. By using the QNRPF, the MPE is enhanced to 0.999 which is close to the ideal value of 1. It indicates that the entropy of the chaotic laser is greatly improved by the QNRPF scheme.

\section{Conclusions}

In conclusion, we develop a experimental technique to generate a high-quality chaotic laser with Gaussian invariant distribution using the QNRPF. The quantum shot noise is extracted through a balanced homodyne detection for building a random phase feedback loop. Meanwhile, a high-speed optical switch is used to reset the laser dynamics to the original state repeatedly and measure the time evolution of the chaotic transient statistical distribution. The effects of various QNRPF bandwidths and phase modulation depths on the transient invariant distributions are investigated experimentally. The results show that the transient intensity distribution of the chaotic output is improved from an asymmetric invariant distribution to a Gaussian invariant distribution. This indicates that the dynamic transition of the chaotic system from microscopic initial noise to macroscopic random fluctuations. Furthermore, as the QNRPF bandwidth and phase modulation depth increase, the TDS of chaos is suppressed to 0.036, approaching the bottom noise floor. The MPE is enhanced to 0.999, which is close to the maximum ideal value of 1. Meanwhile, the skewness of the long-term probability distribution of the chaotic time series has been increased by an order of magnitude. Therefore, the QNRPF provides an effective method for the preparation of high-entropy stochastic sources with the Gaussian invariant distribution, and it would play an important role in high-speed physical random number generation and chaos-based secure communication.

\begin{backmatter}
\bmsection{Funding}
National Natural Science Foundation of China (62175176, 62075154); Key Research and Development Program of Shanxi Province (International Cooperation, 201903D421049).

\bmsection{Disclosures}
The authors declare that there are no conflicts of interest related to this article.

\bmsection{Data Availability Statement}
Data underlying the results presented in this paper may be obtained from the authors upon reasonable request.

\end{backmatter}

%%%%%%%%%%%%%%%%%%%%%%% References %%%%%%%%%%%%%%%%%%%%%%%%%

\end{document}